\documentclass{aa}
\usepackage{txfonts}
\usepackage{rotfloat}
\usepackage{amsfonts}
\usepackage{amsmath}
\usepackage{amssymb}
\usepackage{graphicx}

\usepackage{esint}
\usepackage{color}

\providecommand{\\}{\\}

\usepackage{xcolor}

\makeatother

\usepackage[english]{babel}
\newcommand{\Msun}{\mbox{$M_{\odot}$} }

\newcommand{\eR}{{\rm eROSITA}}

\def\lsim{\;\raise0.3ex\hbox{$<$\kern-0.75em\raise-1.1ex\hbox{$\sim$}}\;}
\def\gsim{\;\raise0.3ex\hbox{$>$\kern-0.75em\raise-1.1ex\hbox{$\sim$}}\;}

\def\kms{\rm ~km~s^{-1}}

\def \kms {\rm ~km~s^{-1}}

\def\arcmin{\hbox{$^\prime$}}
\def\arcsec{\hbox{$^{\prime\prime}$} }

\begin{document}

\title{Spatially resolved X-ray spectra of the galactic SNR G18.95-1.1: SRG/eROSITA view}

\author
{A.M. Bykov\thanks{E-mail: byk@astro.ioffe.ru (AMB)}
\inst{1}
\and
Y.A. Uvarov\thanks{E-mail: uv@astro.ioffe.ru (YAU)}
\inst{1}
\and
E.M. Churazov 
\inst{2,3}
\and
M.R. Gilfanov \thanks{E-mail: mgilfanov@mpa-garching.mpg.de (MRG)}
\inst{2,3}
\and 
P.S. Medvedev 
\inst{2,3}
\\
}

\institute{ Ioffe Institute, 26 Politekhnicheskaya st., St. Petersburg 194021, Russia
\and
Space Research Institute of the Russian Academy of Sciences (IKI), 84/32 Profsoyuznaya st., Moscow 117997, Russia
\and
Max Planck Institute for Astrophysics, Karl-Schwarzschild-Str. 1, D-85741 Garching, Germany \\
}

\date{}


\label{firstpage}

\abstract{}{
We study the X-ray emission of the galactic supernova remnant (SNR) G18.95-1.1 with the \eR~ telescope on board the Spectrum R\"{o}ntgen Gamma ({\it SRG})  orbital observatory. In addition to  the pulsar wind nebula that was previously identified and examined by {\it ASCA} and {\it Chandra}, we study the  X-ray spectra of the bright SNR ridge, which is resolved into a few bright clumps.
}{
The wide field of view and the large collecting area in the $0.2-2.3$ keV energy range of {\it SRG}/{\rm eROSITA} allowed us to perform spatially resolved spectroscopy of G18.95-1.1.
}{
 The X-ray ridge of G18.95-1.1 is asymmetric, indicating either supernova ejecta asymmetry or their interaction with a cloud.  The X-ray dim northern regions outside the pulsar wind nebula can be described by a thin thermal plasma emission with a temperature $\sim$0.3 keV and a solar composition.  The X-ray spectra of a few bright clumps located along the southern ridge may be satisfactorily approximated by a single thermal component of the Si-rich ejecta at the collisional ionization equilibrium with a temperature of about 0.3 keV. The bright ridge can be alternatively fit with a single component that is not dominated by equilibrium ejecta with T$\sim$ 0.6 keV. The high ratio of the derived Si/O abundances indicates that the ejecta originated in deep layers of the progenitor star. The plasma composition of a southern Si-rich clump and the bright ridge are similar to what was earlier found in the Vela shrapnel A and G. 
}{}


\titlerunning{Spatially resolved  X-ray spectra of the Galactic SNR G18.95-1.1: SRG/{\rm eROSITA} view}
\authorrunning{A.M. Bykov et al.}

\maketitle

\section{Introduction}

Supernova remnant (SNR) G18.95-1.1, which is located in the Galactic plane within
the central radian of the Milky Way,  was discovered by \citet{Fuerst_1985Natur}
in the course of the 2.695 GHz Galactic plane survey \citep{Reich_1984A&AS}.
Later on, it was studied at other frequencies in the radio band by 
 \citet{Odegard_1986AJ.....92.1372O}, \citet{Barnes_1988srim.conf}, \citet{Patnaik_1988Natur}, \citet{Furst_1989A&A}, \citet{Fuerst_1997A&A...319..655F}, \citet{Reich_2002nsps.conf....1R}, and  \citet{Sun_2011A&A...536A..83S}.
The radio and X-ray observations revealed that G18.95-1.1 belongs to the class of the
composite-type SNR with a central peak emission. The central emission region
is dominated by a pulsar wind nebulae (PWN) that was studied with the {\it Chandra} X-ray Observatory \citep{Tullman_2010ApJ}. In the optical, the H$_{\alpha}$ imaging of the G18.95-1.1 region was  performed  by \citet{Stupar_2011MNRAS.414.2282S}. 
In X-rays, G18.95-1.1 was observed with the {\it ROSAT} \citep{Aschenbach_1991A&A...246L..32A,Fuerst_1997A&A...319..655F},
{\it ASCA} \citep{Harrus_2004ApJ}, and {\it Chandra} \citep{Tullman_2010ApJ}
orbital X-ray observatories. 
The $\gamma$-ray source FL8Y J1829.5-1254 in the {\it Fermi}-LAT  catalog \citep{Acero_2016ApJS..224....8A} may be associated with this supernova remnant.  

The {\it ROSAT} X-ray spectrum of G18.95-1.1 was fit  by \citet{Aschenbach_1991A&A...246L..32A}  with an absorbed thermal emission model with $N_{H}=3.4\cdot10^{21}$cm$^{-2}$
and $T=0.434$ keV. Their analysis also revealed two other $\chi^2$-statistics minima  (reduced $\chi^2<1$) with parameter values $N_{H}=9.5\cdot10^{21}$cm$^{-2}$,
$T=0.2$ keV, and $N_{H}=5.2\cdot10^{20}$cm$^{-2}$, $T=1.1$ keV, which
where rejected. \citet{Fuerst_1997A&A...319..655F}
obtained a slightly different result with best-fit parameters values $N_{H}=3.4\cdot10^{21}$cm$^{-2}$, $T=0.95$ keV ($\chi^2/dof=0.84$),
and $N_{H}=12.5\cdot10^{21}$cm$^{-2}$, $T=0.25$ keV ($\chi^2/dof=0.86$). They adopted the lower value of
 $N_{H}=(3.4\pm1.5)\cdot10^{21}$cm$^{-2}$, which
is consistent within the 90\% confidence level errors with the $N_{H}=(2.0-2.2)\cdot10^{21}$cm$^{-2}$
obtained from the HI radio observations \citep{Fuerst_1997A&A...319..655F}. On the other hand,  analyzing a joint fit of the {\it ROSAT} and {\it ASCA} spectra, \citet{Harrus_2004ApJ} obtained higher values of $N_{H}$ for a few various plasma emission models: collisional ionization equilibrium
(CIE), the thermal MEKAL (Mewe-Gronenschild-Kaastra) model by \citet{Mewe1985,1993A&AS...97..873K}, and thermal emission with
nonequilibrium ionization (NEI), the PSHOCK model by \citet{Borkowski_2001ApJ...548..820B}. 
The best-fit model parameters were $N_{H}=8.4\cdot10^{21}$cm$^{-2}$,
$T=0.58$ keV for the CIE and $N_{H}=9.4\cdot10^{21}$cm$^{-2}$, $T=0.9$
keV for the NEI models with solar abundances. However, for the single-component models, 
the reduced $\chi^2$  values were rather high ($\chi^2/dof=1.76$, dof=89 for NEI and $\chi^2/dof=3.15$, dof=90 for CIE models), which might be evidence that  more complicated models of the source emission are required that allow abundance variation or additional components. 
The distance estimates to SNR G18.95-1.1 were discussed in detail by \citet{Furst_1989A&A},  \citet{Harrus_2004ApJ}, and  \citet{Tullman_2010ApJ}. It was
suggested that the most plausible distance is $\sim 2$ kpc, which agrees with the
HI measurements by \citet{Furst_1989A&A}. The higher $N_H$ value obtained from the X-ray spectral analysis might be explained when 
molecular $H_2$ or dust are assumed to contribute substantially to absorption in the direction to G18.95-1.1. This is discussed in more detail at the end of section 2.

{\it Chandra} observations of G18.95-1.1 by \citet{Tullman_2010ApJ} were dedicated to study the PWN, and the {\it Chandra} field of view (FOV) only permitted studying its immediate surroundings. 
The temperature of the plasma emission component in the PWN region was found to be $T=0.48$ keV, and the power-law component index was found to vary from 1.4 to 1.9, depending on the
size of the considered emission region. The relatively hard power-law index and the elongated shape of 
the PWN may indicate that the bow shock-type nebula is produced by the interaction with the flow behind the reverse  shock that passed through the PWN, as was proposed for the  Vela PWN by \citet{2011ApJ...740L..26C} and \citet{2017SSRv..207..235B}.

The location of G18.95-1.1 in the inner galactic plane   
makes the spectral analysis rather difficult because of the uncertainty
and inhomogeneity of the background radiation and the contamination by the emission from  point sources  projected at the SNR. The brightest
X-ray point source is the {\it ROSAT} emission excess J182848-130055 discussed
by \citet{Fuerst_1997A&A...319..655F} and \citet{Harrus_2004ApJ}. In the {\it ASCA} map,
this source is located at the position J182849.9-130107.44. \citet{Harrus_2004ApJ}
discussed the possible coincidence of this source with the star J182850.08-130120.3.
Another point source is CXOU J182913.1\textendash{}125113, which is
located at the edge of the PWN region and is a pulsar candidate. However,
it has no radio or optical counterpart \citep{Tullman_2010ApJ}.
There are also other point sources in the vicinity of the SNR. The
{\it Chandra} observatory is ideally suited for identifying and excising the point sources, but its FOV covers only a small part of the G18.95-1.1 remnant. {\it ASCA} and {\it ROSAT} observations covered the whole remnant, but the exclusion of the point sources is more difficult due to their lower angular resolution. 
The nonuniformity  of the background emission and limited {\it ASCA} and {\it ROSAT} FOV areas mean that
obtaining background spectra was difficult as well.

The  extended Roentgen Survey with an Imaging Telescope Array (eROSITA) telescope \citep{Predehl_2020arXiv201003477P} on board  the recently launched  {\it SRG} observatory \citep[][]{2021arXiv210413267S}
provided a good opportunity for studying this remnant. eROSITA has an excellent
sensitivity in the 0.5-2.3 keV energy band, an on-axis angular resolution of $16$ arcsec half-power diameter (HPD) at 1.5 keV,  and a very good  spectral resolution $\sim$ 70 eV full width at half maximum (FWHM) below 1 keV. 
During the performance verification (PV) phase, it observed an area of $\sim25$ square degrees 
in the Galactic plane centered at $l=20\degr$  with a nearly uniform exposure of $\approx7$ ks. 
These observations fully covered  G18.95 -1.1  and its surrounding area. This enables us to accurately measure the background spectrum and to exclude point sources. In this paper, \eR~ data are used to study the SNR G18.95-1.1.

\section{SRG/eROSITA observations of G18.95-1.1}

The $\sim25$ square degree  area of the Galactic ridge around
the latitude $l=20\degr$ was observed by \eR~ in October 2019 during
the PV phase. The region was observed in the raster scan mode, which permitted us to obtain an almost uniform exposure of the scanned region with an effective (vignetting-corrected) exposure of $\approx3.3$ ks (0.5--2.3 keV).
G18.95-1.1  is located toward the edge of the scanned  region. For this reason, the exposure varied across the remnant from $\approx2.1$ ks
in the south to $\approx3.7$ ks in the north. The accurate knowledge of the telescope vignetting \citep{Predehl_2020arXiv201003477P} means that the collected 
data allow detailed imaging and spectral analysis of the SNR. We present this below.

\subsection{Image analysis}

\begin{figure*}
\includegraphics[scale=0.40]{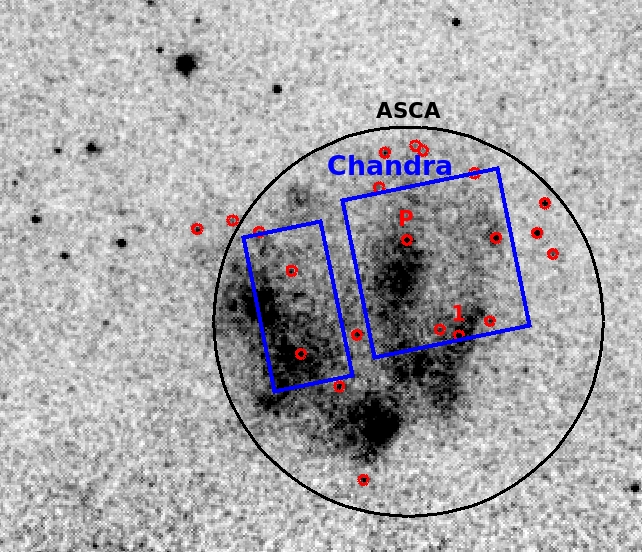} \includegraphics[scale=0.41]{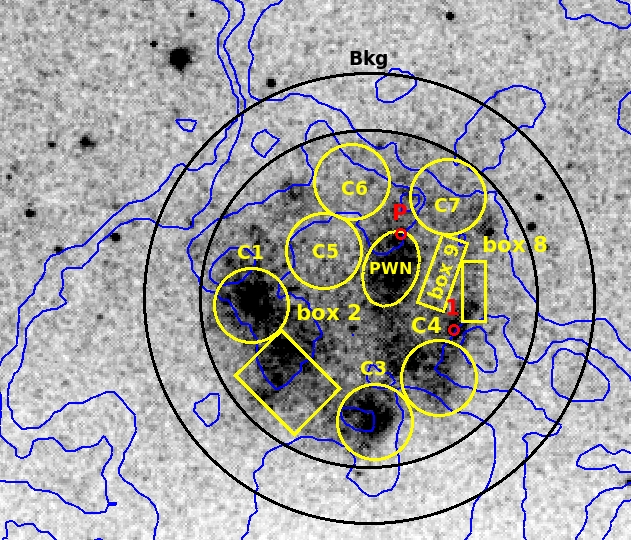}
\caption{\eR~ exposure-corrected map of 
 G18.95-1.1  and its surrounding
area. The {\it ASCA} GIS and {\it Chandra} FOVs are  
shown in the left panel map in black and
blue. Small red circles show the location of point sources that
are excluded from the spectral analysis. The  brightest point source
is marked with a 1, and the pulsar candidate source is marked with a P. $H_{\alpha}$ contours from the SHASSA
survey \citep{Gaustad_2001PASP..113.1326G}, shown as blue curves, are superposed in the right panel map together with the regions we used for the spectrum analysis. The annulus we used to obtain the background spectrum is shown in black. Its inner boundary is the outer
border for the SNR source region. Smaller regions  that were used to extract spectra of various parts of the remnant are shown in yellow, and the brightest point source and pulsar candidate are marked
in red.}
\label{fig:maps1}
\end{figure*}

The \eR~ detectors are most sensitive in the 0.5-2.3 keV energy range.
The X-ray \eR~ map constructed in this energy band is shown in Fig.~\ref{fig:maps1} together with $H_{\alpha}$ contours from
the SHASSA survey \citep{Gaustad_2001PASP..113.1326G}, other mission
FOVs, and the regions we used for the spectral analysis. The $H_{\alpha}$
map is rather complicated in the area surrounding the remnant, but the northeast arc in the $H_{\alpha}$ map, which is very
well correlated with the X-ray SNR shell, and the excess in $H_{\alpha}$
emission correlated with X-ray emission region C3 may be evidence
of the SNR shell $H_{\alpha}$ emission.

\begin{figure}
\includegraphics[bb=40bp 60bp 760bp 660bp,clip,angle=0,scale=0.38]{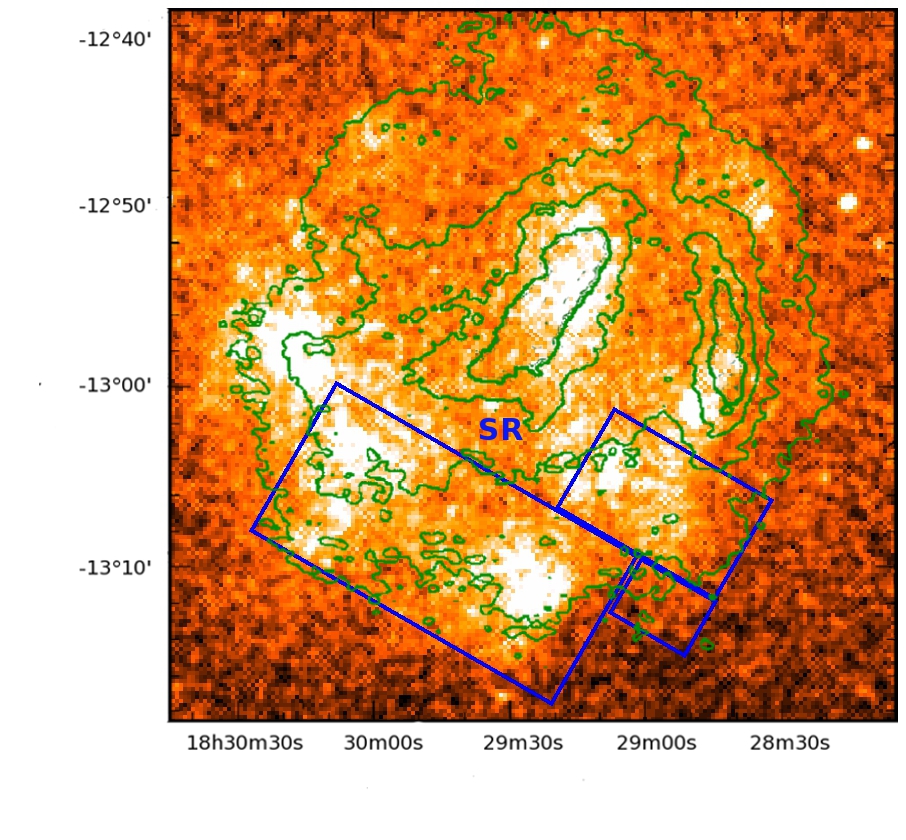}
\caption{\eR~ exposure-corrected map of G18.95-1.1 with superposed green contours of the 10.55 GHz radio emission (contours by E. F{\"u}rst) from \citet{Harrus_2004ApJ}. The composite SR region encompassing regions box2, C3, and C4 is shown by the blue boxes.}
\label{fig:maps2}
\end{figure}

Seven regions in the shell (C1, box2, C3, C4, C6, C7, and box8), the first four of  which coincide with clumps of enhanced X-ray emission, were chosen for the spectral analysis. An SNR region including the whole remnant and the southern ridge (SR) region, encompassing
regions box2, C3, and C4, were also used in the analysis. An elliptical region in the central part of the remnant, which is almost identical to the region $e3$ from  \citet{Tullman_2010ApJ}, was used to study the PWN emission.  These regions are listed in Table~\ref{tab:an_regions} and shown in Figs.~\ref{fig:maps1} and \ref{fig:maps2}. Point sources that are projected onto the source and background regions were excluded from the  analysis. These sources are shown in Fig.~\ref{fig:maps1} and listed in Table \ref{tab:point_sources}.

\begin{table}
\caption{\label{tab:an_regions} Regions we used for the spectral analysis.}
\begin{center}
\begin{tabular}{cccc}
\hline \hline 
region & $\alpha$ & $\delta$ & size/radius, \\
 &  &  & arcmin\\
\hline 
SNR & 18:29:26.7940 & -12:58:03.993 & 18\\
\hline 
Bkg & 18:29:26.7940 & -12:58:03.993 & 18-24\\
\hline 
C1 & 18:30:18.5314 & -12:58:43.695 & 4\\
\hline 
box2 & 18:30:02.635 & -13:06:56.58 & 6.9-8.9\\
\hline 
C3 & 18:29:24.6016 & -13:11:10.213 & 4\\
\hline 
C4 & 18:28:56.2220 & -13:06:27.934 & 4\\
\hline 
C5 & 18:29:46.7551 & -12:52:55.426 & 4\\
\hline
C6 & 18:29:34.5075 & -12:45:33.797 & 4\\
\hline
C7 & 18:28:52.5042 & -12:47:09.432 & 4\\
\hline
box8 & 18:28:41.0472 & -12:57:17.460 & 2.5-6.4\\
\hline
box9 & 18:28:54.8106 & -12:55:16.133 & 2.9-7.8\\
\hline
PWN & 18:29:17.146 & -12:54:52.76 & 2.8-4.1\\
\hline 
\end{tabular}
\end{center}
\tablefoot{J2000 coordinates (RA/DEC) are listed. For the 
background region (Bkg),  the  inner and outer radii of the annulus are given. For the box regions, we give the 
sizes, and for the  elliptical PWN region the semi-axis. The composite SR region shown
in Fig.~\ref{fig:maps2} is not listed in the Table.}
\end{table}

\begin{table}
\caption{\label{tab:point_sources} Point sources excluded from the spectral analysis.}
\begin{center}
\begin{tabular}{cc}
\hline \hline 
$\alpha$ & $\delta$\\
\hline 
18:28:50.040 & -13:01:20.64\\
\hline 
18:29:13.100 & -12:51:13.00\\
\hline 
18:28:15.792  & -12:50:23.64 \\
\hline 
18:28:33.696  & -12:50:56.40\\
\hline 
18:29:31.776  &  -13:16:46.56\\
\hline 
18:28:12.336  & -12:47:15.00\\
\hline 
18:29:34.560 & -13:01:17.76\\
\hline 
18:30:44.520 &  -12:50:00.96\\
\hline 
18:29:22.440 & -12:41:51.72\\
\hline 
18:29:59.064 & -13:03:20.88 \\
\hline 
 18:28:08.928  & -12:52:40.44 \\
 \hline 
\end{tabular}%
\vline
\begin{tabular}{cc}
\hline \hline 
$\alpha$ & $\delta$\\
\hline 
18:30:17.400  & -12:50:20.04 \\
\hline 
 18:29:24.912 & -12:45:34.92 \\
\hline 
18:28:36.384 & -12:59:50.28\\
\hline 
 18:29:42.408  &  -13:06:48.24 \\
\hline 
 18:28:58.416 & -13:00:42.84 \\
\hline 
18:30:29.112 & -12:49:05.16\\
\hline 
 18:30:03.288  & -12:54:26.64\\
\hline 
18:29:05.784  & -12:41:38.04 \\
\hline 
18:29:09.000  & -12:41:05.64 \\
\hline 
18:28:43.248 & -12:44:03.48\\
\hline 
 $\ $ & \\
\hline 
\end{tabular}
\end{center}
\tablefoot{J2000 coordinates (RA/DEC) are listed. The listed sources are marked in  Figure \ref{fig:maps1} by red circles.
Only sources  located
inside the outer border of the  background region are listed.
The radius of the exclusion region is the same for all sources and equals $30\arcsec{}$. 
The two first sources in the left column are the brightest point source (marked with 1 in Fig.~\ref{fig:maps1}) and the pulsar candidate (P in Fig.~\ref{fig:maps1}).}

\end{table}

The position of the {\it SRG} brightest point
source (1) is slightly shifted ($\sim13^{\prime\prime}$) from the position in {\it ASCA} and is almost within the typical $12^{\prime\prime}$ {\it ASCA} 90\% error circle radius.
It also coincides (within  $<1^{\prime\prime}$) with the position of the star J182850.08-130120.3 from 2MASS\footnote{
http://irsa.ipac.caltech.edu/Missions/2mass.html} catalog, which was suggested as the counterpart of the {\it ASCA} brightest point
source by \citet{Harrus_2004ApJ}.

\subsection{Spectral analysis}

The   XSPEC%
\footnote{http://heasarc.gsfc.nasa.gov/xanadu/xspec/%
} package  v. 12.11.1 was used for the spectral analysis \citep{xspec_1996ASPC..101...17A}.
 Thermal emission of optically thin plasma in an SNR is typically described  with  spectral  models either assuming collisional ionization equilibrium or nonequilibrium ionization. \citet{Harrus_2004ApJ} applied both types of models to study the X-ray emission of the G18.95-1.1. The same  model types are used in our analysis as well:  the APEC CIE spectral model, which calculates the emission spectrum of collisionally ionized diffuse gas  using atomic data from the AtomDB%
\footnote{http://http://www.atomdb.org/%
} database, and the PSHOCK NEI model, which is a constant-temperature plane-parallel shocked-plasma emission model \citep[e.g.,][]{Borkowski_2001ApJ...548..820B}. APEC spectral data
v.3.0.9 and eigenfunction data v.3.0.4  were used in XSPEC for the simulations. A TBABS interstellar absorption model with corresponding abundances \citep{Wilms_2000ApJ...542..914W} was used to calculate  the interstellar absorption.

The combined data from \eR~ telescope modules 1-4 and 6, which are unaffected by light leakage \citep{Predehl_2020arXiv201003477P},  were used in the analysis.  Most of the spectra were grouped to have more than 30 counts in a bin with the grppha FTOOLS%
\footnote{http://heasarc.gsfc.nasa.gov/ftools/%
} \citep{FTOOLS_2014ascl.soft08004N} task.  The spectra of the small dim regions box8 and box9 were grouped with more than 10 counts in a bin, and the spectrum of the total SNR was grouped with more than 100 counts in a bin.

\begin{table}
\begin{center}
\caption{\label{tab:one_comp} One-component  CIE (APEC) spectral models  for a low-temperature local $\chi^2$ minimum.}
\begin{tabular}{p{0.8cm}cccc}
\hline \hline
region & $N_{H}$ & $T$ & dof & $\chi^{2}/dof$ \\
 & $10^{22}$ cm$^{-2}$ & keV & &   \\
\hline 
SNR & $1.38_{-0.02}^{+0.03}$ & $0.30_{-0.01}^{+0.01}$ & 315 & $1.77$ \\
\hline 
C1 & $1.17_{-0.06}^{+0.06}$ & $0.30_{-0.02}^{+0.02}$  & 126 & $1.08$ \\
\hline 
box2 & $1.29_{-0.06}^{+0.08}$ & $0.28_{-0.03}^{+0.02}$ & 117 & $1.38$ \\
\hline 
C3 & $1.28_{-0.06}^{+0.07}$ & $0.29_{-0.02}^{+0.02}$ & 93 & $1.03$ \\
\hline 
C4 & $1.08_{-0.08}^{+0.07}$ & $0.31_{-0.02}^{+0.03}$ & 100 & $1.18$ \\
\hline 
C5 & $1.67_{-0.13}^{+0.14}$ & $0.30_{-0.04}^{+0.05}$ & 89 & $1.26$ \\
\hline 
C6 & $1.76_{-0.25}^{+0.26}$ & $0.27_{-0.06}^{+0.07}$ & 38 & $0.97$ \\
\hline 
C7 & $1.51_{-0.18}^{+0.24}$ & $0.39_{-0.09}^{+0.09}$ & 62 & $1.08$ \\
\hline 
box8 & $1.23_{-0.15}^{+0.21}$ & $0.43_{-0.11}^{+0.11}$ & 55 & $1.06$ \\
\hline 
box9 & $1.74_{-0.24}^{+0.25}$ & $0.25_{-0.06}^{+0.07}$ & 84 & $0.86$ \\
\hline 
SR & $1.27_{-0.03}^{+0.05}$ & $0.29_{-0.01}^{+0.02}$ & 227 & $1.45$ \\
\hline 
\end{tabular}
\end{center}

\tablefoot{ All errors are shown with 90\% confidence level. }
\end{table}

\begin{table}
\begin{center}
\caption{\label{tab:one_comp2} One-component  NEI  (PSHOCK) spectral models for a high-temperature local $\chi^2$ minimum.}
\begin{tabular}{p{0.8cm}cccp{0.4cm}p{0.8cm}}
\hline \hline
region &  $N_{H}$ & $T$ & $\tau_{u}$ & dof & $\chi^{2}/dof$\\
 &  $10^{22}$ cm$^{-2}$ & keV & s/cm$^{3}$ & & \\
\hline 
SNR &  $1.08_{-0.04}^{+0.03}$ & $0.62_{-0.03}^{+0.03}$ & $3.6_{-0.7}^{+0.9}\cdot10^{11}$ & 314 & $1.72$\\
\hline 
box2  & $0.92_{-0.08}^{+0.08}$ & $0.73_{-0.09}^{+0.11}$ & $1.9_{-0.7}^{+1.1}\cdot10^{11}$ & 116 & $1.19$\\
\hline 
C3 &  $0.94_{-0.10}^{+0.53}$ & $0.63_{-0.38}^{+0.09}$ & $3.4_{-1.1}^{+1.6}\cdot10^{11}$ & 92  & $1.17$\\
\hline 
C4 & $0.85_{-0.10}^{+0.15}$ & $0.57_{-0.12}^{+0.08}$ & $5.1_{-1.8}^{+3.1}\cdot10^{11}$ & 99 & $1.13$\\
\hline 
C5 & $1.25_{-0.13}^{+0.12}$ & $0.71_{-0.11}^{+0.15}$ & $4.7_{-3.2}^{+6.3}\cdot10^{11}$ & 88 & $1.25$\\
\hline 
SR  & $0.93_{-0.05}^{+0.05}$ & $0.67_{-0.04}^{+0.05}$ & $3.1_{-0.8}^{+1.0}\cdot10^{11}$ & 226 & $1.18$\\
\hline 
\end{tabular}
\end{center}
\tablefoot{Only regions with a high-temperature local $\chi^2$ minimum are listed. All errors are shown with 90\% confidence level.}

\end{table}

The results of the spectral fitting with single-component models are presented in Tables \ref{tab:one_comp} and \ref{tab:one_comp2} for the set of regions shown in Fig.\ref{fig:maps1}. Earlier, these models  were used by  \citet{Harrus_2004ApJ}, who showed that  single-component model fitting typically yields two local minima in $\chi^{2}$ : at  low ($\sim0.3$)
keV  and high ($\sim0.6-0.9$ keV) plasma temperatures.   We obtained a similar result  for
the spectra of the southern regions (box2, C3, C4, and SR) and of the whole SNR, but the spectra of the northern regions (C6, C7, box8, and box9) only allow low-temperature fits.  

The spectra of the northern regions C6, C7, box8, box9, and region C1 have appropriate model fits ($\chi^2/dof < 1.1$)  with the APEC model and solar abundances \citep{Wilms_2000ApJ...542..914W}.
In contrast, the spectra of the whole SNR, box2, and SR regions provide a far poorer agreement with  the single-temperature component CIE or NEI fits and solar abundances. They provide a $\chi^2/dof \approx$1.4 for the SR region and a  $\chi^2/dof \approx$ 1.7 for the whole SNR. 

\begin{figure*}
\includegraphics[bb=0bp 20bp 702bp 550bp,clip,angle=0,scale=0.36]{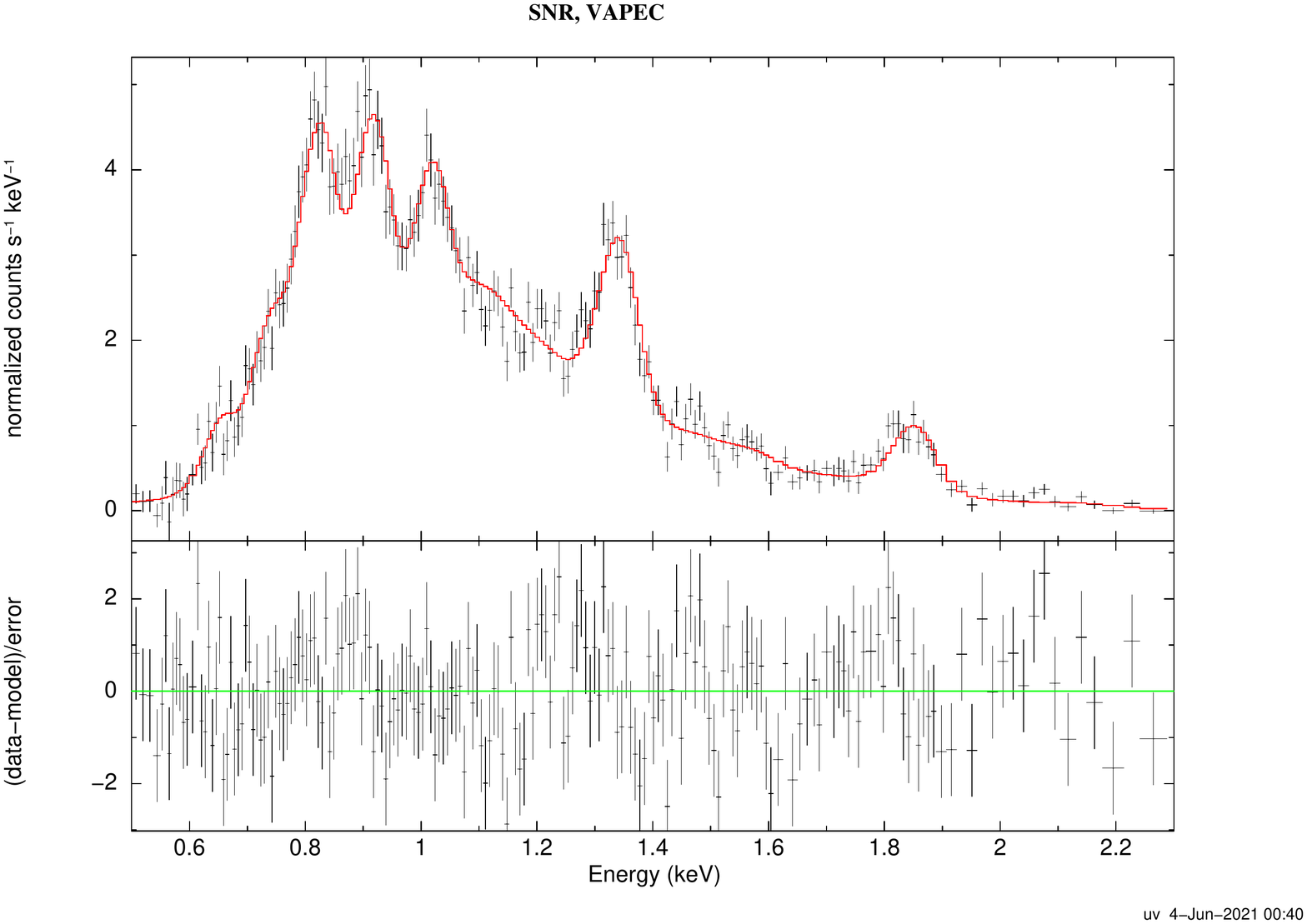}\includegraphics[bb=0bp 20bp 702bp 550bp,clip,angle=0,scale=0.36]{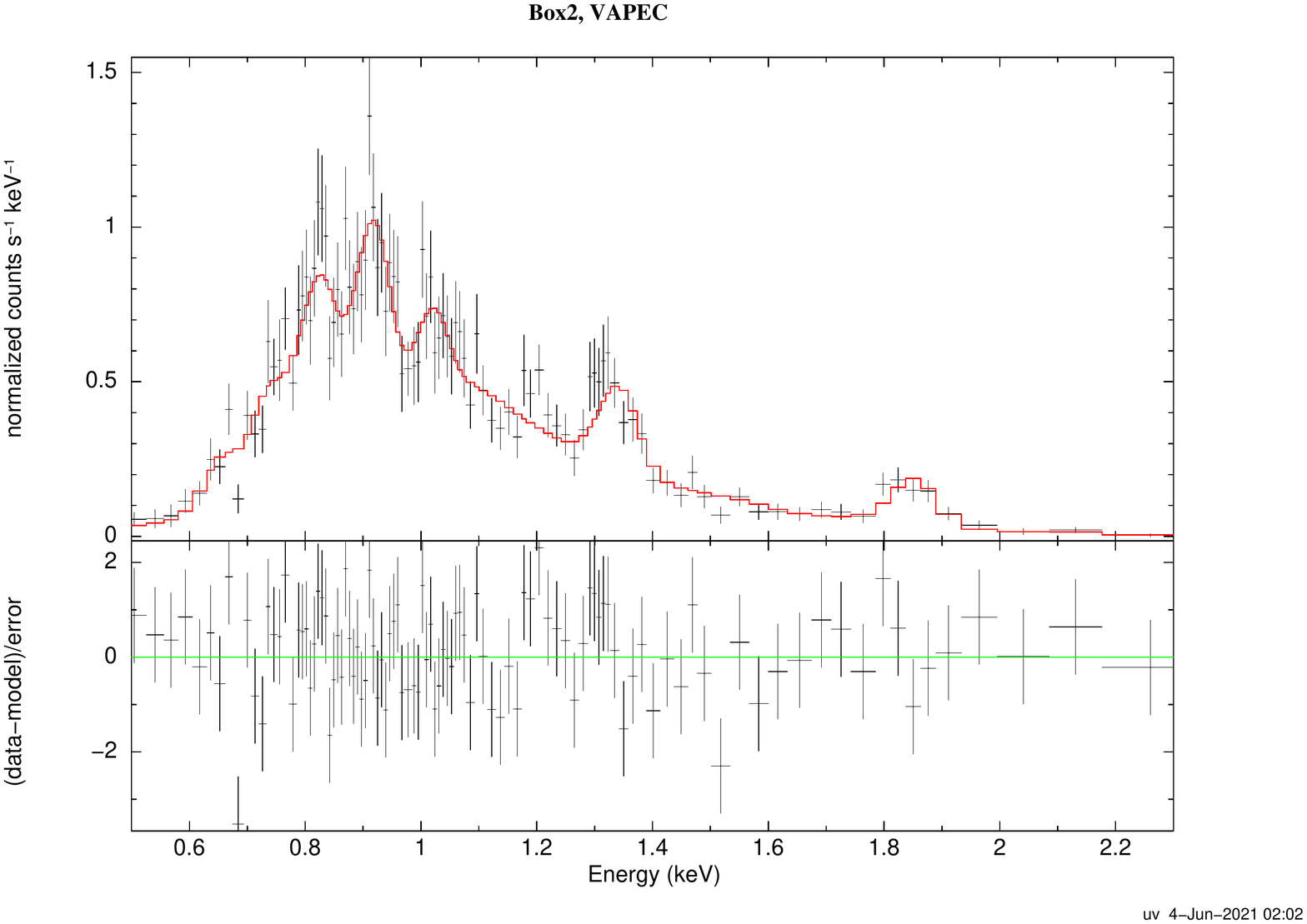}

\includegraphics[bb=0bp 20bp 702bp 550bp,clip,angle=0,scale=0.36]{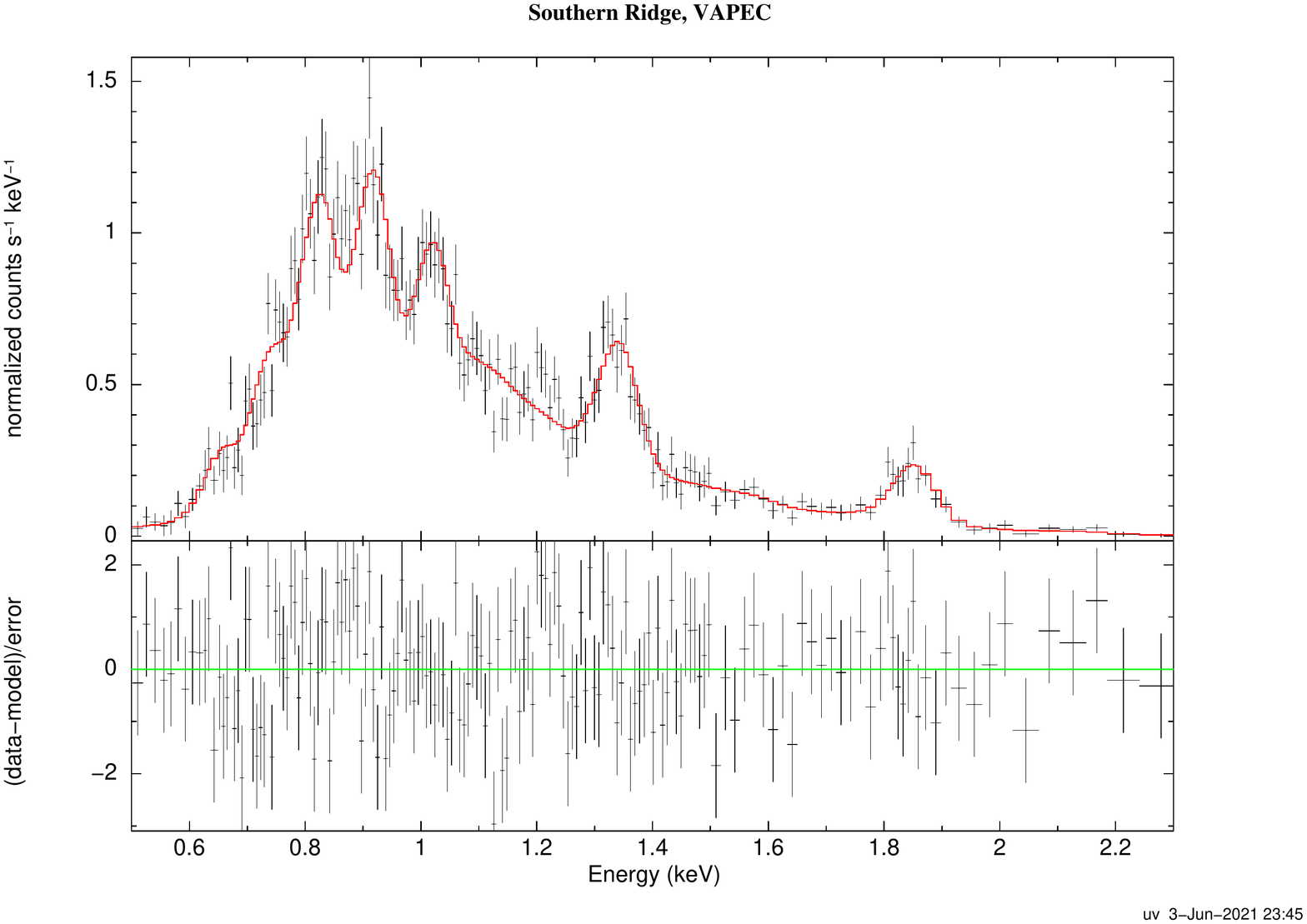}\includegraphics[bb=0bp 20bp 702bp 550bp,clip,angle=0,scale=0.36]{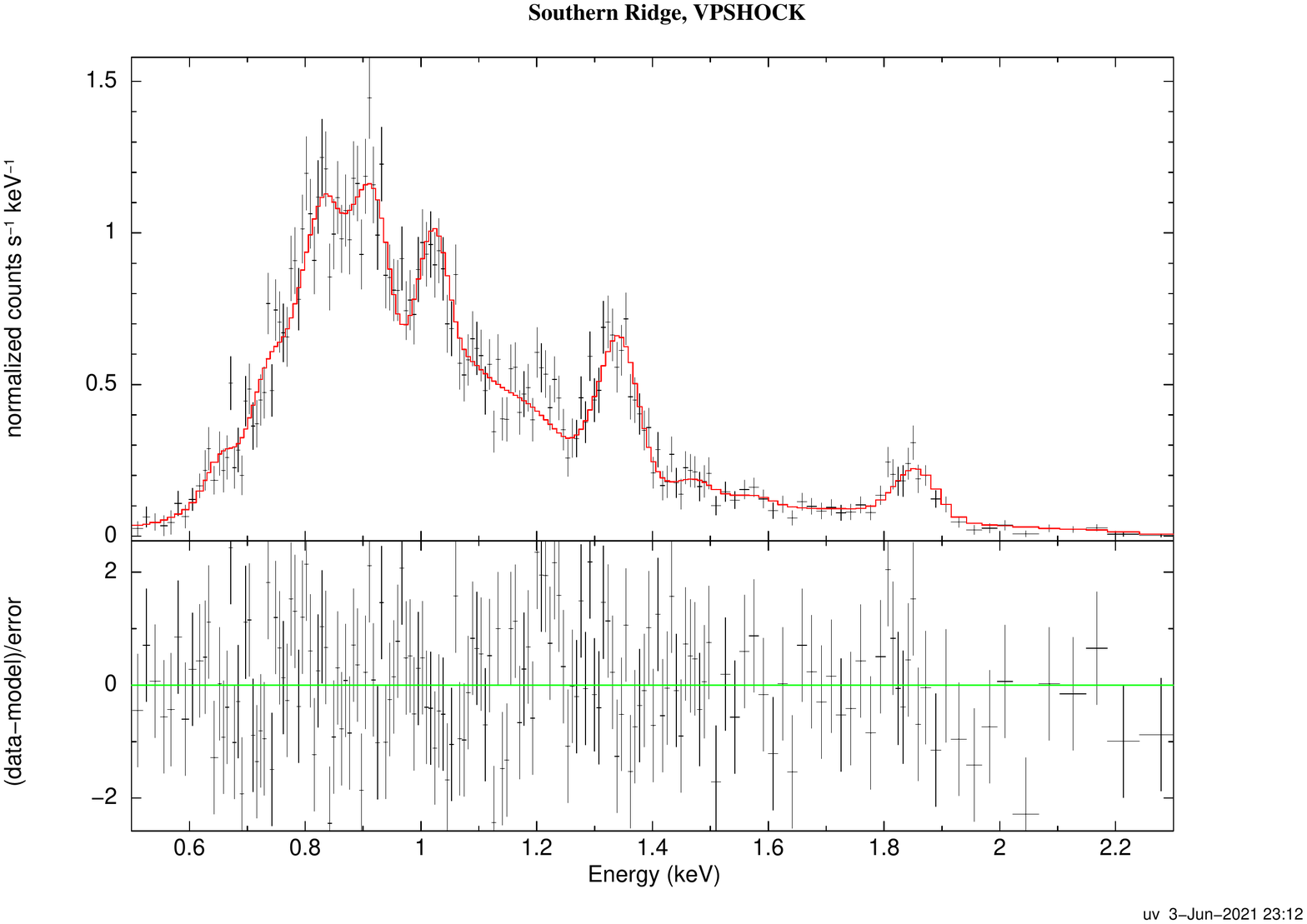}

\caption{Spectra of the whole SNR, box2, and SR regions fit with spectrum models with variable abundances of Ne, Mg, Si, and Fe (the Ni abundance was assumed equal to Fe). The upper left panel shows the spectrum of the whole SNR with the VAPEC  model,
the upper right panel shows the spectrum of the box2 region with the VAPEC  model,
 the lower left panel shows the spectrum of the SR region with the VAPEC model, and the lower right panel shows the spectrum of the SR region with the high-temperature VPSHOCK model. Each panel shows the spectrum data (black) and model (red) in the upper graph and the residuals in the lower graph. The model parameters for these spectra are given in Table~\ref{tab:tab_abund}. }
\label{fig:spec_abun}
\end{figure*}

Varying the element abundances enables a significant improvement of the fits. We obtain a $\chi^2/dof\leq 1.03$ for box2 and the SR regions and a $\chi^2/dof\approx$1.2 for the SNR. The fitting results are shown in Table~\ref{tab:tab_abund}. The abundances of Ne, Mg, Si, Fe, and Ni were varied (Ni and Fe abundances were assumed to be equal)\footnote{Here and below, the element abundances are measured relative to solar abundances.}.
  Si is significantly overabundant, while the overabundance of Ne and Mg is moderate and is compatible with the solar value with a 90\% confidence level  for the spectrum of  the box2 region. The Fe (and Ni) abundance is typically below the solar value. The discussed spectra are shown in Fig.~\ref{fig:spec_abun}.
The spectrum of the SR region allows 
approximation by both low- and  high-temperature
models with similar values of the $\chi^2/dof \approx$ 1.03. The spectra of the box2 region and of the whole remnant allow only low-temperature fits ($\chi^2/dof=1.0$ for
box2 and $\chi^2/dof=1.17$ for the SNR).

For the low-temperature NEI models, the ionization timescale $\tau_u$ is consistent with longest timescales covered by NEI models in XSPEC , and it is unconstrained on the upper side. The typical lower limits on $\tau_u$ are comparable to or higher than $\sim10^{12}\,{\rm s\,cm^{-3}}$  for most regions. In this case, the use of CIE APEC (or VAPEC) models is justified and appropriate. These CIE models are listed in Table~\ref{tab:one_comp}.
The lowest allowed values of $\tau_u$ ($\sim4\cdot10^{11}\,{\rm s\,cm^{-3}}$) are found for regions C1, box8, and C7. The best-fitting values of $\tau_u$ for these three regions are  $9.0\cdot 10^{11}$, $3.9\cdot 10^{12}$, and $\gsim 4\cdot 10^{13}\,{\rm s\,cm^{-3}}$, respectively. While the value of $\tau_u$ is unconstrained on the upper side for these regions as well, it cannot be excluded that NEI effects play a role in these spectra. More observations are needed to clarify this point.
For the higher-temperature PSHOCK (VPSHOCK) models, we obtain $\tau_u\sim3\cdot10^{11}$~s/cm$^{-3}$ (Table \ref{tab:one_comp2}), therefore it is preferable to use  NEI models in this case. Below, we therefore used
CIE and NEI models to describe  low- and high-temperature models, respectively.

Multitemperature spectral models can also improve spectral fits for the regions of box2 and the SNR in comparison with the single-component models with fixed solar abundances. Table \ref{tab:multi_comp} presents the examples of spectral fitting results for the two-component models. The derived temperatures generally agree, but are somewhat below those obtained by \citet{Harrus_2004ApJ} from the combined ROSAT PSPC and ASCA GIS analysis 
of the SNR. The multicomponent models can be justified in case of the extended SNR G18.95-1.1 emission. The single-component models with varied abundances
allow describing emission spectra with a lower $\chi^2/dof$ ratio for localized regions that can be associated with the stellar ejecta.

\begin{table*}
\caption{\label{tab:tab_abund} One-component CIE (VAPEC) and NEI (VPSHOCK) spectral models with variable abundances.}
\begin{center}
\begin{tabular}{ccccc}
\hline \hline
 model     & SR          & SR      & box2 & SNR\\
 parameters        & $tbabs*vpshock$         & $tbabs*vapec$       & $tbabs*vapec$ & $tbabs*vapec$\\
\hline 
$N_{H}, $ $10^{22}$cm$^{-2}$        &$0.74_{-0.09}^{+0.09}$  & $1.00_{-0.09}^{+0.09}$              &  $0.92_{-0.14}^{+0.15}$ & $1.07_{-0.06}^{+0.08}$ \\
T, keV         & $0.59_{-0.08}^{+0.07}$          & $0.31_{-0.02}^{+0.02}$         & $0.30_{-0.04}^{+0.04}$ & $0.31_{-0.01}^{+0.02}$\\

Ne             & $2.3_{-0.9}^{+1.6}$             & $1.5_{-0.5}^{+0.9}$            & $1.8_{-0.8}^{+4.2}$    & $1.1_{-0.3}^{+0.5}$\\
Mg             & $2.7_{-1.1}^{+2.0}$             & $1.7_{-0.6}^{+1.5}$            & $2.2_{-1.3}^{+6.3}$    & $1.6_{-0.5}^{+0.8}$\\
Si             & $3.6_{-1.8}^{+3.8}$            & $5.4_{-2.3}^{+5.0}$            & $8.5_{-5.5}^{+28.5}$ & $3.9_{-1.3}^{+2.3}$\\
Fe             & $1.4_{-0.5}^{+0.9}$              & $0.9_{-0.3}^{+0.6}$            & $1.0_{-0.5}^{+2.0}$ & $0.8_{-0.2}^{+0.3}$\\
dof            & 234                         & 235                            & 122 & 332\\
$\chi^{2}/dof$ & 1.01                               & 1.01                           & 1.00 & 1.17\\
$\tau_{u}$,  s/cm$^{3}$     & $4.1_{-1.1}^{+1.6}\cdot10^{11}$  &  --                             & -- & --\\
Norm, $\frac{10^{-14}}{4\pi D^2}\int n_{e}n_{H}dV$       & $6.3_{-2.8}^{+5.2}\cdot10^{-3}$  & $4.0_{-1.9}^{+3.2}\cdot10^{-2}$ & $6.7_{-5.2}^{+11.3}\cdot10^{-3}$ & $1.1_{-0.4}^{+0.6}\cdot10^{-1}$ \\
\hline 
\end{tabular}
\end{center}
\tablefoot{Models for SR, box2 and the whole SNR
regions. All errors are shown with 90\% confidence level.
$n_{e}$ and $n_{H}$ are electron and H densities (cm$^{-3}$), and $D$ is the distance to the source.}

\end{table*}

\begin{table*}
\caption{\label{tab:multi_comp} Two-component spectral models.}
\begin{center}
\begin{tabular}{ccccccccc}
\hline  \hline 
 & $N_{H}$ & $T_{1}$ & $N_{1}$ & $T_{2}$ & $N_{2}$ & $\tau_{u}$ & dof & $\chi^{2}/dof$\\
region & $10^{22}$ cm$^{-2}$ & keV & $\frac{10^{-14}}{4\pi D^2}\int n_{e}n_{H}dV$  & keV & $\frac{10^{-14}}{4\pi D^2}\int n_{e}n_{H}dV$  & s/cm$^{3}$ &  & \\
\hline 
SNR   & $1.33_{-0.03}^{+0.04}$ & $0.19_{-0.01}^{+0.01}$ & $0.50_{-0.09}^{+0.14}$ & $0.47_{-0.02}^{+0.02}$ & $7.4_{-0.7}^{+1.0}\cdot10^{-2}$ & - & $313$ & $1.33$\\
\hline 
box2 & $1.10_{-0.12}^{+0.11}$ & $0.24_{-0.05}^{+0.05}$ & $1.6_{-1.0}^{+2.4}\cdot10^{-2}$ & $0.76_{-0.18}^{+0.46}$ & $2.5_{-1.3}^{+2.1}\cdot10^{-3}$ & $3.0_{-2.2}^{+5.0}\cdot10^{11}$ & $114$ & $1.09$\\
\hline 
PWN\tablefootmark{a}  & $1.66_{-0.13}^{+0.10}$ & $0.26_{-0.03}^{+0.02}$ & $4.1_{-1.7}^{+2.5}\cdot10^{-2}$ & - & - & - & $106$ & $0.91$\\
\hline 
\end{tabular}
\end{center}
\tablefoot{Models tbabs*(apec+pshock), tbabs*(apec+apec), and tbabs*(apec+pow) for box2, the whole SNR,  and PWN regions, respectively. All errors are shown with 90\% confidence level. $n_{e}$ and $n_{H}$ are electron and H densities (cm$^{-3}$), and $D$ is the distance to the source.
\\
\tablefoottext{a}{The power-law component of the PWN region spectral model has index $3.1_{-1.2}^{+1.1}$
and a normalization $6.1_{-4.0}^{+12.0}\cdot10^{-4}$ ph/keV/cm$^{2}$/s at 1 keV.}
 }

\end{table*}

\begin{figure}
\includegraphics[bb=0bp 20bp 702bp 550bp,clip,angle=0,scale=0.35]{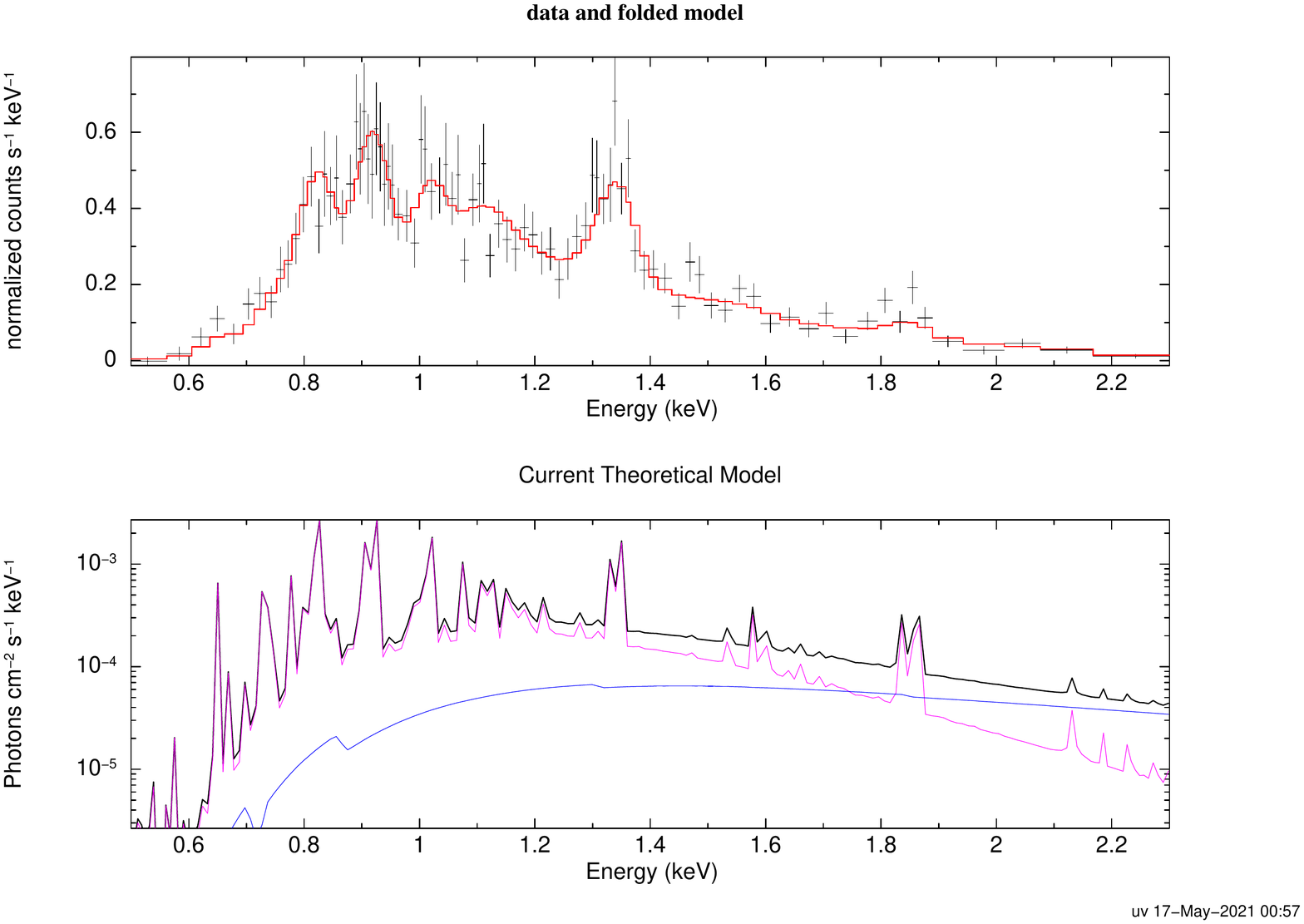}
\caption{Spectrum of the PWN region. We show 
spectrum data (black) and the model (red)
in the upper panel and the model composition in the lower panel. The entire
model is shown with the black curve, and the low-temperature apec component is shown
with the magenta curve. The blue curve shows the power-law component. The model spectrum parameters are listed in 
Table~\ref{tab:multi_comp}.  }

\label{fig:spectra}
\end{figure}

Despite the rather good model approximations obtained with VAPEC (VPSHOCK) models, there are 
 notable residuals in the spectra of the whole SNR, box2, and SR regions. They are shown in Fig.\ref{fig:spec_abun}. 
 The residuals have several characteristic peaks at the energies, suggesting that they are caused mostly by the emission lines of Fe, Si,  and possibly Ne, which are imperfectly reproduced by the simple one-component model used above (Fig. \ref{fig:spec_abun}). The excess at $\sim1.2-1.3$ keV is most interesting here because it indicates a connection with the Fe L emission line complex. This requires further investigation.

For the  spectral analysis of the emission from the PWN region, which is almost identical to region $e3$ from \citet{Tullman_2010ApJ},  a two-component spectral model of $tbabs*(apec+power\: law)$ was used. 
The model parameters for this fit are listed in Table \ref{tab:multi_comp} and its spectrum is shown in Fig. \ref{fig:spectra}. The power-law index agrees within  the statistical errors with
the results of \citet{Tullman_2010ApJ}, but the plasma temperature $T$ appears to be somewhat lower, and the $N_{H}$ value appears to be somewhat higher than that from {\it Chandra} analysis. The power-law model-component uncertainties are larger than the uncertainties reported by \citet{Tullman_2010ApJ} because the high-energy response of \eR\ is limited. The spectral model $tbabs*(vapec+power\: law)$ with variable abundances (Ne, Mg, Si, Fe, and Ni) does not improve the fit for this region much ($\chi^2/dof=0.83$), but suggests fewer restrictions on $N_H=1.43^{+0.22}_{-0.25}\cdot10^{21}$~cm$^{-2}$, $T=0.28\pm0.04$~keV, and $\Gamma=2.0_{-1.8}^{+2.7}$ that is, it
corresponds better with the results of other nearby region fitting and with the  $N_{H}$ estimation by \citet{Tullman_2010ApJ}. Varied abundances have large statistical errors and are consistent with
 solar  for all  elements that we varied (within 1~$\sigma$ for Ne, Mg, and Si and within 90\% confidence level for Fe), while the best-fit values indicate a possible overabundance of Si (1.6) and lack of Fe (0.4).

For  the whole SR region, the VPSHOCK and VAPEC spectral models have  $N_{H}$ values of $(0.75\pm 0.1)\cdot10^{22}$~cm$^{-2}$ and  $(1.0\pm0.1)\cdot10^{22}$~cm$^{-2}$, respectively.  These values are  significantly higher than $N_{H}=(3.4\pm1.5)\cdot10^{21}$ cm$^{-2}$ obtained from the {\it ROSAT} data analysis in \citet{Aschenbach_1991A&A...246L..32A}, \citet{Fuerst_1997A&A...319..655F}, and $N_{H}=(2.0-2.2)\cdot10^{21}$ cm$^{-2}$ obtained from the HI radio observations in \citet{Fuerst_1997A&A...319..655F}, but they are close to the $N_{H}$ values discussed in the {\it ASCA} observation analysis in \citet{Harrus_2004ApJ} and obtained from the PWN spectrum analysis based on {\it Chandra} data in \citet{Tullman_2010ApJ}. 

\begin{figure}
\includegraphics[bb=75bp 20bp 400bp 310bp,clip,scale=0.8]{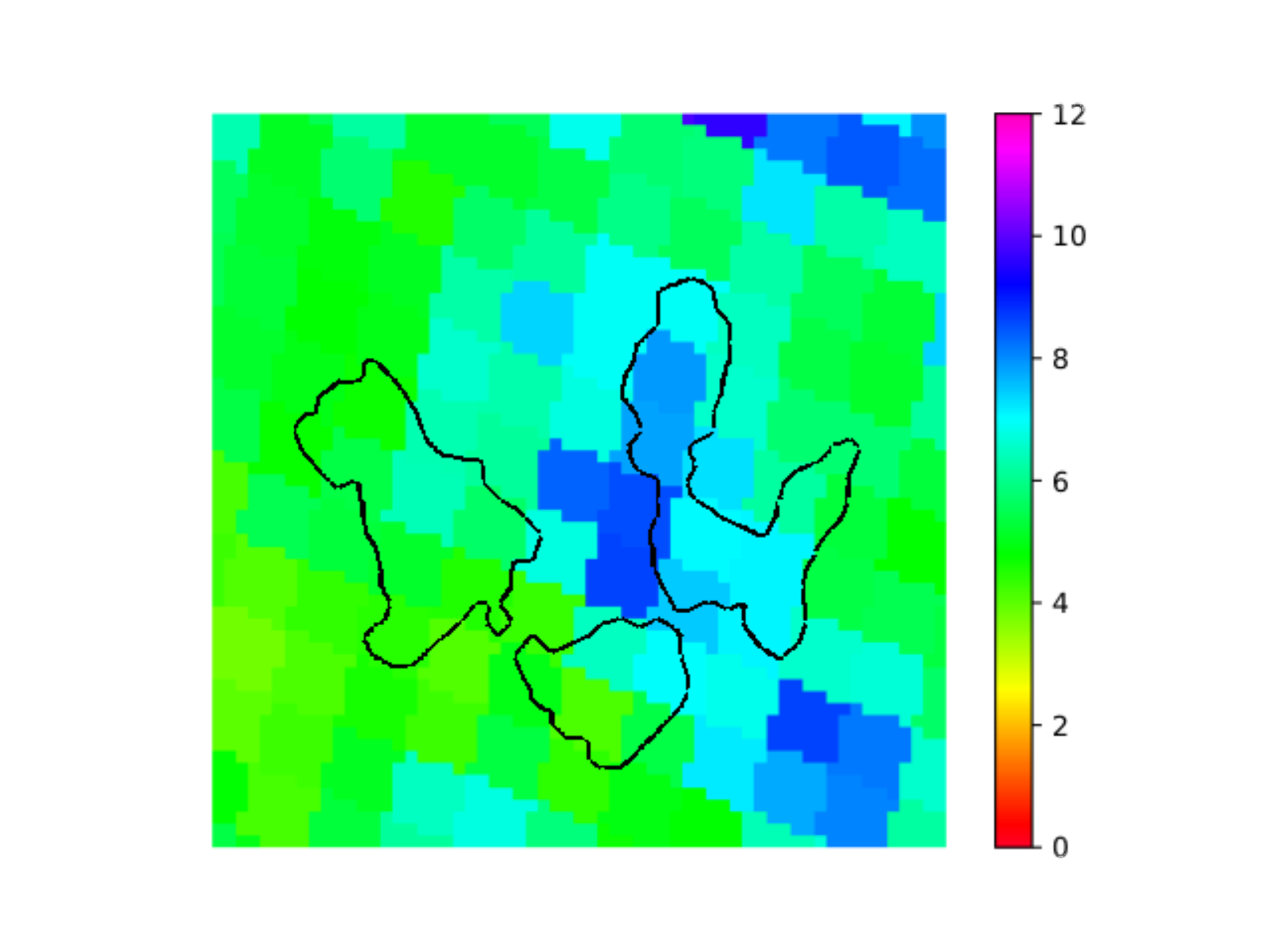}
\vfill
\includegraphics[bb=75bp 20bp 400bp 310bp,clip,scale=0.8]{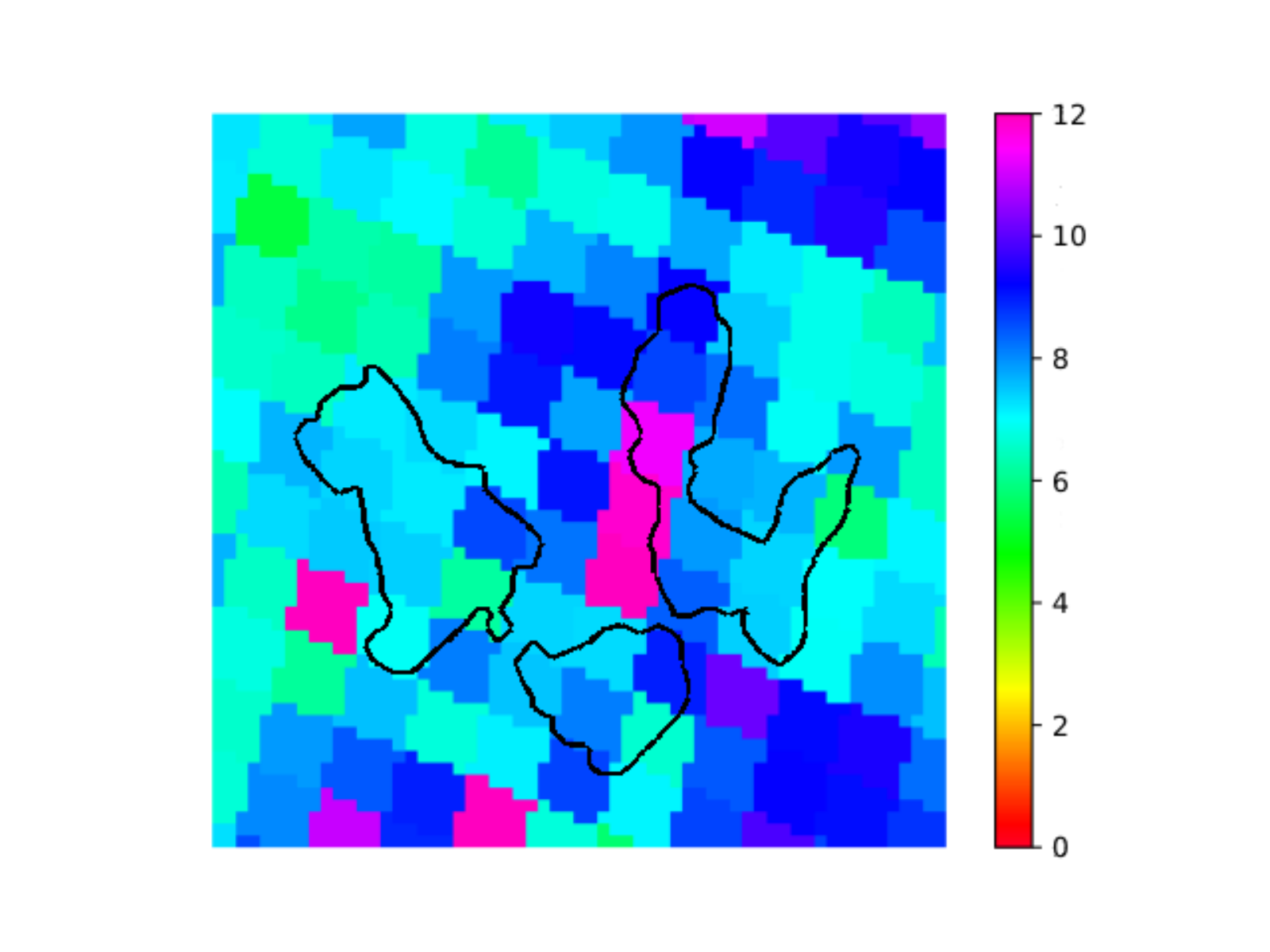}
\caption{Hydrogen column density maps in the direction to SNR G18.95-1.1 derived from the {\sl bayestar19} data cubes \citep{bayestar2019ApJ...887...93G} using the DUSTMAPS Python package \citep{dust2018JOSS....3..695M} for the two distances of 2 kpc (upper panel) and of 3 kpc (bottom panel). The contours of the X-ray emission as detected by {\it SRG} \eR~ in Fig. \ref{fig:maps1} are overlaid in both panels. The color-map bars
show the estimated $N_H$ values in units of $10^{21}$ cm$^{-2}$.}  
\label{fig:bay19}
\end{figure}

We analyzed dust-reddening {\sl bayestar19} data \citep{bayestar2019ApJ...887...93G} in the direction toward SNR G18.95-1.1 using the DUSTMAPS\footnote{http://dustmaps.readthedocs.io/en/latest/}
Python package \citep{dust2018JOSS....3..695M}. In Fig.\ref{fig:bay19} we present the maps of $N_H$ that were derived from the {\sl bayestar19} data cubes using the 
relation between $N_H$ and reddening $N_H\approx8.9\cdot10^{21}E(B-V)$~cm$^{-2}$ from  \citet{Foight2016ApJ...826...66F} for two distances under investigation,  $2$ and $3$ kpc. 
It is interesting that the $N_H$ values derived from X-ray spectroscopy and the {\sl bayestar19} data would agree if the distance to SNR G18.95-1.1 were $\sim$ 3 kpc, which is only slightly larger than suggested earlier. More observational data are needed to refine these estimates and potentially improve the accuracy of the SNR G18.95-1.1 distance determination.

\section{Discussion}
SNR G18.95-1.1 has a complex asymmetric morphology.  The bright emission regions shown in Fig. \ref{fig:maps1} are located within a ring with an inner radius of about 9\arcmin ~ and an outer radius of about 15\arcmin.   The $H_{\alpha}$ half-shell emission excess of the same width observed in the southeast part of the remnant \citep[SHASSA survey,][]{Gaustad_2001PASP..113.1326G}
coincides with the X-ray ridge (see Fig. \ref{fig:maps1}).

The maximum X-ray surface brightness of the structures corresponds to the angular distance of about 12\arcmin ~  from the apparent center of the SNR. Therefore the outer radius of the the bright ridge is within a ring 7.5 pc \lsim R \lsim 12 pc at 3  kpc distance. 

The structure of the X-ray ridge in Fig. \ref{fig:maps1} has a clear large-scale asymmetry that is most prominent in the southeastern part of the SNR. The optical H$_\alpha$ emission in Fig.~\ref{fig:maps1} (bottom panel) is more homogeneous in the field of G18.95-1.1. The X-ray ridge is not apparent in the radio images at 1.4 and 4.9 GHz presented by \cite{Furst_1989A&A}, where one bright arc-like structure is associated with the PWN, while the second bright structure is located well away from the X-ray ridge.  It is difficult to search for the X-ray synchrotron emission filaments produced by the ultrarelatevistic electrons accelerated at the supernova blast wave, which are more prominent in the 4-6 keV energy band,  in the \eR~ observations with $\sim3$ ks exposure, keeping in mind that it is most sensitive below 2.3 keV. The radio maps do not show any thin bright filaments that might be associated with a forward shock \citep{Fuerst_1997A&A...319..655F}, however.

Core-collapse supernovae are in many cases expected to expand into a wind bubble produced by a massive progenitor star  \citep[see, e.g.,][and the references therein]{Chevalier89,Slane_CRab_like00,Dwarkadas2005ApJ...630..892D,Chevalier_Fransson17}. The lack of a clear signature of the forward shock (or outer shells) in the northern part of the G18.95-1.1 \eR~ and radio images can be understood if the asymmetric supernova ejecta is expanding in the diluted matter of the wind cavern. The broad clumpy southern X-ray ridge structure may be a result of the gas heating by the reverse shock that passed the ejecta  \citep[see, e.g.,][]{Chevalier_Fransson17,Vink17,Raymond_2018SSRv..214...28R}. 

\citet{2013A&A...559A..69G} performed 2D-axisymmetric hydrodynamical simulations of the evolution of the circumstellar medium shaped by stellar winds of rotating massive stars at solar metallicity for a large grid of stellar models from 15 to 120 \Msun. The simulations made for different models of the massive star environments with different mass-loss events: the main-sequence wind, the red-supergiant  wind and Wolf-Rayet wind demonstrated the position of the fast wind termination shock at a different distances $\sim$ 5 pc where a density jump is located. In particular, 
a 15 \Msun mass star of 15 Myr age blow a wind  cavern of radius $\sim$ 10 pc  filled with hot gas of a temperature $\sim$ 0.5 keV. The size of the cavern and the gas temperature are increasing with the mass of the star. A progenitor star of mass      $15 \leq M \geq 20$ \Msun with a cavern filled with a hot gas of $\sim$ keV temperature and of a size $\gsim$ 10 pc could provide a viable scenario to explain the appearance of G18.95-1.1 as observed by \eR.
 In the scenarios of Type Ib/II supernovae produced by a core collapse of a massive star, the ejected material may collide with the density jump of the wind termination shock thus forming the observed X-ray ridge. The lack of a clear signature of the forward shock (and the outer shells) in the northern part of G18.95-1.1 both the \eR~ and radio images may be understood if the asymmetric supernova ejecta is expanding in the diluted matter of the wind cavern.     

The ejecta are clumped because of hydrodynamical instabilities. The apparent north-south asymmetry of the emitting clumps can be a result of the recoil of the ejecta in the southern direction, just in the opposite direction to the apparent proper motion of the pulsar, which produces the observed elongated bow-shock-like PWN. The asymmetry of ejecta in Cas A and some other SNRs observed in the optical, radio, and X-ray bands by \citet{2001ApJS..133..161Fesen}, \citet{2015Sci...347..526M}, \citet{Slaneea15}, \citet{CasA_LOFAR18}, and \citet{CasA_ejecta21} was explained as due to the NS kick in the  asymmetric core collapse  \citep[see, e.g.,][]{asymmetry17,CasA_ejecta_Laming20}. Recently, asymmetric X-ray structures were studied with {\it Chandra} in shocked ejecta of core-collapsed SNR  G320.4-1.2/MSH 15-52, where Ne-Mg rich ejecta were found \citep[][]{Borkowski20}.       

The ionization timescale  parameter in Table~\ref{tab:one_comp2} for  the clumps of the SR derived with the PSHOCK spectral model of the temperature T $\sim$ 0.6 keV  is consistent with the characteristic age of the shocked plasma of 4000 to 6000 years,  which was earlier estimated by  \citet{Furst_1989A&A},  \citet{Harrus_2004ApJ}, and  \citet{Tullman_2010ApJ}. The spectral model with T$\sim$ 0.3 keV requires a minimum ionization timescale of $\sim$ 10$^{12}$~s ${\rm \,cm^{-3}}$ , which would rather imply an older age of $\gsim 10^4$ years.

The single-component spectral models presented 
in Table~\ref{tab:tab_abund} clearly favor the high overabundance of Si compared to the solar composition in the SR regions. The NEI VPSHOCK model for the SR region have statistically acceptable fits for metal-rich ejecta with somewhat different temperatures of about 0.3 and 0.6 keV, respectively  (and ionization timescales $\gsim 10^{12}$ and $4.1_{-1.1}^{+1.6}\cdot10^{11}$ s/cm$^{3}$), while the
CIE VAPEC model allows only the low-temperature solution. 
The low-temperature NEI VPSHOCK fit (not shown in
Table~\ref{tab:tab_abund}) requires a high value of the ionization timescale, so that the
CIE VAPEC model is  appropriate in this case.
To study the composition of the SR and box2 regions, we fixed the abundances of C,N,O at the solar values because the available count statistics and relatively high $N_H$ values estimated for  G18.95-1.1 do not allow us to obtain meaningful estimates of the C,N,O abundances. We found that a good CIE VAPEC model fit with a fixed O abundance provided Si/0 and Si/Fe of $\sim$ 5-8  and Si/Mg and Si/Ne of $\sim$ 3-4. We recall the nearby Vela SNR \citep[see, e.g.,][]{Vela_shrapnelA_Chandra01}, where the situation was more favorable for studies of abundance variations, including that of the oxygen.

The X-ray image of the Vela SNR obtained with {\sl ROSAT} by \citet[][]{Vela_shrapnel95} discovered  six extended X-ray features outside the forward shock. It was proposed that these features are fast-moving ejecta fragments formed by instabilities during the collapse of the progenitor star. Two of the fragments, dubbed shrapnel A and G, are located at the northeastern and southwestern edges of the Vela SNR, respectively. Using a single-temperature NEI model with T $\sim$ 0.5 keV, \citet[][]{Vela_Si_rich_Katsuda06} estimated the abundances in shrapnel A to be  O $\sim$  0.3, Ne $\sim$ 0.9, Mg $\sim$ 0.8, Si $\sim$ 3, and Fe $\sim$ 0.8 relative to their solar values. A similar Si-rich composition with relatively weak abundances of O, Ne, Mg, and Fe was derived with the two-temperature CIE and single-temperature NEI models from {\sl XMM-Newton} spectra of shrapnel G \citet[][]{Vela_Si_rich17}. Moreover, the Si-group-illuminated jets or pistons in Cas A are very prominent in X-ray, optical,
and infrared emission \citep[see, e.g.,][]{DeLaney2010ApJ...725.2038D,LopezFesen18}.  A possible way of  formation of the bright rings of Si/S-rich material was demonstrated in the modeling of asymmetric ejecta of Cas A by \citet[][]{anisotr_ejecta_orlando16}.  
The high ratios of Si/O$\sim$ 5-10 derived in shrapnels A and G in the Vela SNR and in the bright southern clumps in G18.95-1.1 may be understood if the observed ejecta material came from the deep inner layers of the progenitor star.

 The X-ray emitting mass in  G18.95-1.1 can be estimated from the apparent angular sizes of the SR clumps given in Table~\ref{tab:an_regions} and the X-ray flux normalization factors derived in the single-component spectral models listed in Table~\ref{tab:tab_abund}.
  Assuming some simplified geometry of the clumps and given the uncertainties mentioned above, we can estimate the mass of X-ray emitting plasma as a few solar masses. Most of the emission from the SR region comes from the cores of three clumps (C1, box2, and C3). All of them have comparable values of the $\tau_u$ parameter of $\sim 3\times 10^{11}$ s cm$^{-3}$  derived in the Table \ref{tab:one_comp2} and radii about 2 arcmin. Assuming the age of G18.95-1.1 to be younger than 10,000 yr and a  distance of 3 kpc, we can estimate the X-ray emitting clump masses to be just above one solar mass, which is consistent with the mass estimates obtained from the spectrum normalization given in Table \ref{tab:tab_abund}.  Deep multiwavelength  observations of the SNR G18.95-1.1 ridge are required to reduce the uncertainties in the spectral model  and obtain more accurate mass estimation.
 
 There are some known uncertainties \citep[see, e.g.,][]{Greco20} in supernova ejecta mass estimations from the plasma emission measure due to the degeneracy between the derived best-fit values of element abundances (namely a possibility of helium-rich ejecta, which is hard to constrain from X-rays). Therefore it is difficult to estimate the metal ejecta mass. Only the total X-ray emitting mass and the relative abundances of some elements can be estimated at the current level. 
  The presence of the PWN \citep[see][]{Tullman_2010ApJ} and the ejecta mass of a few \Msun estimated above are consistent with a type Ib or IIb SN from a moderately massive progenitor star with an initial mass below 20 \Msun.

We found in the spectra of the clumps (box2, C1, and C3), the SR, and the whole SNR G18.95-1.1 
marginally significant spectral features at photon energies in the range 1.2 - 1.3 keV.  If real, these spectral features could be attributed to the emission of Fe XVII - Fe XX from the Fe L complex \citep{Gu2019A&A...627A..51G}, or as an alternative explanation,  the 1.2-1.3 keV residuals could be the emission of Ne X Ly$_{\beta}$, Ly$_{\gamma}$, and Ly$_{\delta}$ lines
\citep{Cumbee2016MNRAS.458.3554C}. 

The Fe-rich ejecta knots with the prominent Fe L complexes were found in type Ia  LMC SNRs DEM L238 and DEM L249 \citet{Fe_ejecta_Borkow06} and the RGS {\sl XMM-Newton} spectra of the type Ia supernova remnant Tycho (SN 1572).  \citet{Williams2020ApJ...898L..51W}. Some "blobs" with a significant 1.25 keV line emission were found also in a mixed-morphology SNR G350.0-2.0
by \citet{Karpova2016MNRAS.462.3845K}. 
Therefore, more observations are
needed to confirm or reject the apparent spectral excess in SNR G18.95-1.1.

\section{Summary}
The X-ray \eR~ image of G18.95-1.1 revealed a complex asymmetric structure with a bright ridge of emission located mainly in the SE part of the remnant and a bright  radially elongated structure. This structure was found in the radio imaging, and {\it Chandra} observations later confirmed that it is most likely a PWN. The apparent position and shape of the bow-shock-type PWN indicate a pulsar proper velocity of a few hundred $\kms$. The asymmetric shape of the X-ray ridge may be understood as the result of a recoil of the material that is ejected after the core collapse and given the direction of apparent motion of the pulsar due to its initial kick.               

The wide  FOV of \eR~ and the scanning observation mode provided a fairly uniform exposure across the SNR. This  allowed us to study different  background models. The column density values   $N_{H} \sim  (7.5-10)\cdot10^{21}$~cm$^{-2}$  derived here from the spectra of the clump regions and the whole SNR generally agree with that obtained  from the {\it ASCA} data analysis by \citet{Harrus_2004ApJ} and {\it Chandra} studies of the PWN region by \citet{Tullman_2010ApJ}. These $N_{H}$ values are  greater than those that were estimated earlier in a {\it ROSAT} data analysis  \citep[][]{Aschenbach_1991A&A...246L..32A,Fuerst_1997A&A...319..655F} and HI radio observations by \citet{Furst_1989A&A}. The high $N_{H}$ values obtained in the X-ray data analysis of {\it ASCA}, {\it Chandra,} and \eR~ suggest that the distance to G18.95-1.1 is about 3 kpc, as illustrated in Fig.~\ref{fig:bay19}, while the issue requires further study. 

The good spectral resolution of \eR~  revealed a double-peaked spectral structure just below 1~keV, where two spectral features of Fe-L and Ne lines at energies about 0.8 and 0.9 keV are clearly separated. 
In the \eR~  data analysis of G18.95-1.1, we have modeled
the X-ray  emission from the spatially resolved structures with both the collisional ionization equilibrium  and nonequilibrium ionization XSPEC models. The single thermal CIE  model with variable abundances provides  satisfactory fits for both the dim northern region and the X-ray bright SR with a temperature about 0.3-0.4 keV. However, while the northern regions with temperatures $\sim 0.4$ keV allow for a solar composition, the bright southern regions require a strong silicon overabundance with a lower temperature $\sim 0.3$ keV. Alternatively, the southern regions can be fitted with a single-temperature NEI model with a temperature $\sim 0.6$ keV and a strong Si overabundance as well. The Si-rich clumps in G18.95-1.1  are similar to the ejecta shrapnel A and G discovered in the Vela SNR.   
 The X-ray morphology and spectra of G18.95-1.1 detected with \eR~ can be understood in the scenario of a core-collapse supernova with Si-rich ejecta fragments that expanded into the wind of the massive progenitor star.                    
\section{Acknowledgements}
This work is based on observations with eROSITA telescope onboard {\it SRG} observatory. The {\it SRG} observatory was built by Roskosmos in the interests of the Russian Academy of Sciences represented by its Space Research Institute (IKI) in the framework of the Russian Federal Space Program, with the participation of the Deutsches Zentrum für Luft- und Raumfahrt (DLR). The {\it SRG}/eROSITA X-ray telescope was built by a consortium of German Institutes led by MPE, and supported by DLR.  The {\it SRG} spacecraft was designed, built, launched and is operated by the Lavochkin Association and its subcontractors. The science data are downlinked via the Deep Space Network Antennae in Bear Lakes, Ussurijsk, and Baykonur, funded by Roskosmos. The eROSITA data used in this work were processed using the eSASS software system developed by the German eROSITA consortium and proprietary data reduction and analysis software developed by the Russian eROSITA Consortium.

The authors thank the anonymous referee for a careful reading of the paper and helpful comments which we used to improve the data analysis and interpretation. 
The authors thank R.A. Sunyaev for a helpful comment. 
A.M.B. and Yu.A.U. were supported by the RSF grant 21-72-20020. Some of the modeling was performed at the Joint Supercomputer Center (JSCC) RAS and at the ``Tornado'' subsystem of the St.~Petersburg Polytechnic University supercomputing center.
\bibliographystyle{aa}
\bibliography{paper}

\end{document}